\documentclass[3p,times]{elsarticle}

\usepackage{ecrc}

\volume{00}

\firstpage{1}

\journalname{Submitted to Astronomy and Computing}

\runauth{Vaduvescu, Curelaru and Popescu}

\jid{A&C}

\jnltitlelogo{Astronomy and Computing}

\usepackage{amssymb}
\usepackage{listings}
\usepackage{comment}
\usepackage{paralist}
\usepackage{longtable}

\usepackage[figuresright]{rotating}

\begin{document}

\begin{frontmatter}



\dochead{Submitted to Astronomy and Computing (25 Mar 2019)}

\title{Mega-Archive and the EURONEAR Tools for Datamining \\ World Astronomical Images}


\author{Ovidiu Vaduvescu}
\ead{ovidiu.vaduvescu@gmail.com}
\address{Isaac Newton Group (ING), Apt. de correos 321, E-38700, Santa Cruz de La Palma, Canary Islands, Spain \\ 
         Instituto de Astrofisica de Canarias (IAC), Via Lactea, 38205 La Laguna, Tenerife, Spain}

\author{Lucian Curelaru}
\address{Amateur astronomer and computer programmer, Brasov, Romania}

\author{Marcel Popescu}
\address{Instituto de Astrofisica de Canarias (IAC), Via Lactea, 38205 La Laguna, Tenerife, Spain \\
         Universidad de La Laguna, 38205 La Laguna, Tenerife, Spain}

\begin{abstract}
The world astronomical image archives represent huge opportunities to time-domain astronomy 
sciences and other hot topics such as space defense, and astronomical observatories should 
improve this wealth and make it more accessible in the big data era. 
In 2010 we introduced the {\it Mega-Archive} database and the {\it Mega-Precovery} server for 
data mining images containing Solar system bodies, with focus on near Earth asteroids (NEAs). 
This paper presents the improvements and introduces some new related data mining tools 
developed during the last five years. 
Currently, the {\it Mega-Archive} has indexed 15 million images available from six major 
collections (CADC, ESO, ING, LCOGT, NVO and SMOKA) and other instrument archives and surveys. 
This meta-data index collection is daily updated (since 2014) by a crawler which performs 
automated query of five major collections.
Since 2016, these data mining tools run to the new dedicated EURONEAR server, and the 
database migrated to SQL engine which supports robust and fast queries. 
To constrain the area to search moving or fixed objects in images taken by large mosaic cameras, 
we built the graphical tools {\it FindCCD} and {\it FindCCD for Fixed Objects} which overlay the 
targets across one of seven mosaic cameras (Subaru-SuprimeCam, VST-OmegaCam, INT-WFC, VISTA-VIRCAM, 
CFHT-MegaCam, Blanco-DECam and Subaru-HSC), also plotting the uncertainty ellipse for poorly 
observed NEAs. 
In 2017 we improved {\it Mega-Precovery}, which offers now two options for calculus of the 
ephemerides and three options for the input (objects defined by designation, orbit or observations). 
Additionally, we developed {\it Mega-Archive for Fixed Objects} (MASFO) and {\it Mega-Archive 
Search for Double Stars} (MASDS). 
We believe that the huge potential of science imaging archives is still insufficiently exploited. 
In this sense, defining and making available a standard format for indexing meta-data needed to 
access the image archives could strongly enhance their use. We recommend to IAU to define such a
standard and ask the astronomical observatories to adopt it for indexing their image archives in
a homogeneous manner, and make these indexes available up to date, free of any proprietorship period. 
\end{abstract}

\begin{keyword}
Data mining \sep asteroids \sep near Earth asteroids (NEAs) \sep image archives \sep Mega-Archive \sep Mega-Precovery. 

\end{keyword}

\end{frontmatter}



\section{Introduction}
\label{sec1}

\noindent
The world astronomical image archives provide valuable means to improve the physical properties 
of Solar System bodies, and in particular of near Earth asteroids (NEAs) which remain observable 
for short period of times. NEAs represent laboratories for studying the formation and evolution 
of the minor planets and their physical interactions with Sun and the major planets. 
Part of NEAs, potentially hazardous asteroids (PHAs) and virtual impactors (VIs) could pose some 
risk due to their possibility of impact, but they also represent an opportunity for cheaper 
space missions and eventually future mining industries. \\

Upon discovery, the recovery and follow-up of NEAs are essential for providing the initial 
orbital solution and for searching of possible linkage with previously known objects. In 
most cases, smaller NEAs fade rapidly and become invisible even for largest telescopes which
are expensive to access and usually lack time for urgent reaction. However, the existing 
image archives represent a free opportunity to improve the orbital knowledge based on 
serendipitous encounters of targets searched using dedicated data mining tools. \\

Searching for fixed objects (like stars or galaxies) in image archives is straightforward, 
because only the position of the target is needed to compare with the known telescope pointing 
and instrument field. The largest astronomical observatories or their collaborating institutions 
provide simple web searching tools or 
more sophisticated services which allow searches of fixed objects in their image archives. 
In 2009 P.~Erwin released the TELARCHIVE Python code\footnote{http://www.mpe.mpg.de/~erwin/code} 
(which requires Linux installation) allowing searches of fixed objects by querying a few remote 
collections of image archives. Data mining of moving objects and especially those having less 
accurate orbits becomes more complex, requiring the intersection in space and time of the orbit 
with the searched archives. \\

Thanks to their large field, few major photographic plate archives started to be used two decades 
ago in the first NEA data mining projects by 
D. Steel in Australia (AANEAS)\footnote{http://users.tpg.com.au/users/tps-seti/spacegd4.html}, 
A. Boattini in Italy (ANEOPP, \citep{Boattini2001}) and 
G. Hahn in Germany (DANEOPS)\footnote{https://web.archive.org/web/20171221091544/http://earn.dlr.de/daneops}. 
Due to their initial tiny sizes, CCD cameras have been less appealing to the astronomical community 
for data mining. Only few major NEA surveys such as NEAT and Spacewatch (whose archives were integrated 
in SkyMorph\footnote{https://skys.gsfc.nasa.gov/cgi-bin/skymorph/mobs.pl}) or the modern Pan-STARRS 
(not providing a public tool) allow precovery searches of known asteroids and NEAs in their own archives. \\

Within the EURONEAR\footnote{http://www.euronear.org} project, since 2007 we datamined some major 
image archives to improve known NEA orbits, involving many amateurs and students in a few public outreach 
and educational projects. In 2010 we published the {\it Precovery} tool which allows searches of all know 
NEAs in a few existing and any other given instrument archive indexed in a simple ASCII meta-data format. 
Using this tool with four powerful archives (CFHTLS-MegaCam, ESO/MPG-WFI, INT-WFC and Subaru-SuprimeCam) 
we improved the orbits of more than 400 NEAs searched through 800,000 images 
\citep{Vaduvescu2009,Vaduvescu2011,Vaduvescu2012,Vaduvescu2013,Vaduvescu2017}. \\

Similar work has been carried out recently for detection and data mining of asteroids in the Kilo-Degree 
Survey (KiDS) observed with the VST-OmegaCam \citep{Mahlke2018} which resulted in 20,221 candidate objects 
(about half known and unknown asteroids) detected in 346 sq.deg. 
Other authors proposed similar public asteroid data mining and citizen science projects. In 2012, S.~Gwyn 
proposed the ``MegaCam Archival Asteroid Search Verification'' (MAASV) project focusing on the CFHT-MegaCam 
archive \citep{Gwyn2012c}. Following our call for collaboration in 2010, the former EURONEAR member 
E.~Solano proposed the SVO-NEA \citep{Solano2014} citizen science tool to precover NEAs in the SDSS DR8, 
then later the similar SVO-ast project to measure NEAs and Mars-crossers in a collection of surveys 
(SDSS, UKIDSS, VISTA and VSS) \citep{Solano2018}. \\

Besides astrometry, image archives are valuable also for deriving photometry and physical properties of 
astronomical objects. Popescu et al. \citep{Popescu2016,Popescu2018} retrieved the asteroids imaged by 
VISTA Hemisphere Survey, obtaining near-infrared colors for 53,447 solar system objects including 57 NEAs, 
431 Mars Crossers, 612 Hungaria asteroids, 51382 main-belt asteroids, 218 Cybele asteroids, 267 Hilda 
asteroids, 434 Trojans and 29 Kuiper Belt objects. \\

In 2010 we introduced the {\it Mega-Precovery} project \citep{Popescu2010}, aiming to allow public searches 
of one or a few asteroids or NEAs in a large collection of image archives which aimed initially to include 
at least one million images. This project's capabilities and databases have been strengthened during the last 
years \citep{Char2013,Popescu2014,Vaduvescu2013} and in this paper we present the latest version of 
{\it Mega-Precovery}. 
To our best knowledge, there is only another similar web-tool allowing searches of asteroids in a large 
collection of archive images, namely the ``Solar System Object Image Search'' 
(SSOIS)\footnote{http://www.cadc-ccda.hia-iha.nrc-cnrc.gc.ca/en/ssois} hosted at CADC in Canada. This 
project was published first in 2012 by S.~D.~J.~Gwyn and colleagues \citep{Gwyn2012a,Gwyn2012b}, being 
focused first only on the CADC archive and enhanced later with many other archives
\citep{Gwyn2014,Gwyn2015a,Gwyn2015b,Gwyn2016}. 
Another similar project to search for moving objects in image archives was ``The Planetary Archive'', 
which was announced in 2014 \citep{Penteado2014} but whose outcome is unknown.

\section{Mega-Archive}
The largest astronomical observatories and related data centres provide {\it science archive 
collections} taken by their instruments and telescopes. These archives are served by common user 
interfaces allowing anybody to search and download images, spectra, catalogs and other data products.
Following our first EURONEAR {\it Precovery} projects \citep{Vaduvescu2009,Vaduvescu2011,Vaduvescu2013}, 
in 2010 we joined the first three {\it instrument archives} (CFHTLS, ESO/MPG and ING/INT) to start 
building the {\it Mega-Archive} database \citep{Popescu2010}, with the aim to index at least one 
million images and add more archives later. 

\subsection{Archive Format and Index}
\vspace{+1em}
To search for any fixed or moving objects, any archive collection must contain some basic 
information about the observed images. To index the instrument archives within our database, 
the following ``meta-data'' fields are essential to be included: \\

\begin{compactitem}
  \item {\bf Image ID} - \emph{the string needed for downloading the FITS image file}
  \item {\bf Observing date and time} - \emph{start of exposure in JD format (at least 5 decimals)}
  \item {\bf Telescope pointings} - \emph{J2000 equatorial coordinates RA (hours with decimals) and DEC (degrees with decimals)}
  \item {\bf Exposure time} - \emph{in seconds}
  \item {\bf Filter} - \emph{given as a string}
  \item {\bf Targeted object or field} - \emph{the name of the actual object or the observed field (string)}
\end{compactitem} 
\vspace{+1em}

The actual FITS images are linked remotely to the servers which store the archive collections, and they 
come either in raw original state or sometimes processed (ex from surveys). To avoid any missing in the 
indexing process of archives, and to avoid any duplicates due to possible processed images, the 
{\it Mega-Archive} includes only the raw images from any given archive. In the future we plan to add 
in the output list links to the processed images, whenever these are available. \\

The instrumental archives are summarized in the master ASCII index file \texttt{ArchiveLogs.txt}, 
which describes all instruments with the following information (one instrument in one line): \\

\begin{compactitem}
  \item {\bf Collection/Telescope-Instrument} - \emph{Three acronyms naming the instrument archive}
  \item {\bf Web Address} - \emph{Root internet address serving a given current image ID}
  \item {\bf FOV} - \emph{Field of view on sky (in sq.arcmin) of the instrument assumed a rectangle}
  \item {\bf MPC} - \emph{Minor Planet Centre observatory code where the instrument is located}
  \item {\bf Width} - \emph{Width of the field (in degrees) along RA}
  \item {\bf Height} - \emph{Height of the field (in degrees) along DEC}
  \item {\bf Mag} - \emph{Limiting visual magnitude (V-band) in one minute exposure, 
        taking into account the telescope diameter\footnote{We adopt the following conventions: 
        V=19 for 0.4m, V=21 for 1m, V=23 for 2m, V=25 for 4m, V=26 for 8m class telescopes and HST}}
  \item {\bf JD\_START} - \emph{Julian date of the first image in the archive}
  \item {\bf JD\_END} - \emph{Julian date of the last image in the archive}
  \item {\bf Nr. imgs} - \emph{Number of images in the archive indexed in the given period}
\end{compactitem}
\vspace{+1em}

\subsection{Instrument Etendue and Archive Etendue}
\vspace{+1em}

To characterize the efficiency of telescopes and instruments in survey work, the astronomers use the 
term {\it etendue} ($A\Omega$) defined as the product of the telescope collecting area $A$ (expressed in 
square meters) and the surface on sky of the imaging camera $\Omega$ (in square degrees). 
To characterize the data mining efficiency of entire instrument archives, we propose the term 
{\it archive etendue} ($A\Omega\,A$) defined as the product of the etendue and the number of science 
images included in the given instrument archive.

\subsection{Archive Collections}
\vspace{+1em}

We briefly present the six image archive collections included in {\it Mega-Archive} by 23 Feb 2019. \\

{\bf The Canadian Astronomical Data Centre (CADC)} was established in 1986 to store Hubble Space Telescope 
(HST) and later Canada France Hawaii Telescope (CFHT) data. Today it manages data collections taken with 
many other North American telescopes. Since the beginning of our project we ingested some of the CFHT archives. 
Currently, the {\it Mega-Archive} includes about two million images from 29 instruments (visible and NIR) 
installed on 8 telescopes linked to the CADC collection. 
The CADC Advanced Search\footnote{http://www.cadc-ccda.hia-iha.nrc-cnrc.gc.ca/en/search} allows 
multiple archive selections using a CGI form which presents a few selection menus. The CADC archive 
also allows programmatic queries, and for each instrument we used the following options to retrieve 
the needed meta-data for all available optical and NIR images (all other search options being left 
as default): 

\emph{
\begin{compactitem}
   \item Obs. Constraints = SCIENCE DATA ONLY 
   \item Band = OPTICAL + INFRARED
   \item Cal. Level = RAW STANDARD
   \item Data Type = IMAGE
   \item Obs. Type = IMAGING + OBJECT
   \item Instrument = Cycling one by one
   \item Temporal Constraints = Obs Date eventually split in intervals to avoid 30,000 lines limit
   \item Results page = Download CSV
\end{compactitem}
}

\vspace{+1em}

{\bf The European Southern Observatory (ESO)} started in 1994 to archive NTT images \citep{Albrecht1994}, 
and provides today one of the largest collection taken with major ESO telescopes in La Silla and Paranal. 
The {\it Mega-Archive} includes more than 3.3 million images from 15 instruments installed on 9 ESO 
telescopes. 
In order to query the ESO archive, we used the {\it ESO Science Archive 
Facility}\footnote{http://archive.eso.org/eso/eso\_archive\_main.html}. It allows to search and 
retrieve Raw Data using a CGI form showing two main selection blocks about the target/program and 
observing information. For each instrument we used the following options to search all available images 
(all other existing search options being left as default): 

\emph{
\begin{compactitem}
   \item Return = ALL FIELDS
   \item Category = SCIENCE
   \item Type = OBJECT 
   \item Mode = IMAGE
   \item Imaging = Cycling instruments one by one
   \item Start/End = Eventually split in intervals to avoid browser crashes
   \item Output preferences = ASCII CSV/DOWNLOAD
\end{compactitem}
}

\vspace{+1em}

{\bf Las Cumbres Observatory Global Telescope (LCOGT)} provides since 2013 their Science Archive 
\citep{Lister2013}, storing all the images taken with the instruments and telescopes of this network. 
In 2017 we collected 21 available imagers from all telescopes, treating each instrument mounted on 
any telescope as one distinct archive (due to the distinct locations and MPC codes of each telescope). 
Today {\it Mega-Archive} includes more than one million images from the LCOGT collection. 
The LCOGT Science Archive\footnote{https://archive.lco.global} could be manually or programatically 
queried based on the following options: 

\emph{
\begin{compactitem}
   \item ObsType = EXPOSE
   \item Reduction Level = RAW
   \item Site = Cycling all
   \item Telescope = Cycling all
   \item Showing = 1000 records per page, then cycling all pages
\end{compactitem}
}

\vspace{+1em}

{\bf The Isaac Newton Group (ING)} built in 2014 the ING science data archive based on 
telescope observing logs. We used this facility to ingest in the  {\it Mega-Archive} almost 
1.5 million images taken with 8 instruments mounted at all 3 ING telescopes. 
Besides the night observing logs, the ING query 
form\footnote{http://www.ing.iac.es/astronomy/observing/inglogs.php} allows an ``Advanced search 
and browse'' section for retrieval of meta-data and links to the available ING images sorted based on 
a few constraints. We used the following parameters to search all available ING images (all other 
available search options on this form being left as default): 

\emph{
\begin{compactitem}
   \item Telescope = ALL
   \item Mode = IMAGING
   \item Frame Type = SCIENCE 
   \item Dates Start/End = Eventually split in intervals to avoid 10,000 lines limit
   \item Output format = ASCII
\end{compactitem}
}

\vspace{+1em}

{\bf The U.S. National Optical Astronomical Observatory (NOAO)} provides since 2007 the 
{\it NOAO Science Archive} (known also as the NOAO NVO Portal) \citep{Miller2007} which 
allows VO-compliant access to imaging data taken with major American telescopes in Arizona 
and Chile. 
In 2012 we manually selected the first NVO instrument archives to ingest in {\it Mega-Archive}.
Until today we included almost 5 million NVO images from 19 imaging instruments mounted on 
10 telescopes. 
The NVO collection is available for searches from the NOAO Science Archive 
website\footnote{http://archive.noao.edu}. Two options are possible, proprietary data access 
(requiring login) and general search for NOAO data (available for any other user). We used 
the second choice with the following options in the Simple Query Form: 

\emph{
\begin{compactitem}
   \item Telescope and Instrument = Cycling all imagers one by one
   \item Observing Calendar Date = Eventually split in intervals to avoid 50,000 lines limit
\end{compactitem}
}

\vspace{+1em}

{\bf The Subaru-Mitaka-Okayama-Kiso Archive (SMOKA)} was built in 2001 to store the images 
taken with major Japanese telescopes \citep{baba2002}. It requires user registration and login 
before providing the FITS images, but the meta-data could be retrieved without registration. 
The {\it Mega-Archive} includes more than one million images from 17 instruments installed on 
4 Japanese telescopes. 
The SMOKA Archive Advance Search\footnote{https://smoka.nao.ac.jp/fssearch.jsp} provides 
meta-data selected based on the instruments and date intervals, allowing an output of maximum 
20,000 rows in one query. We used the following parameters to search all available SMOKA images: 

\emph{
\begin{compactitem}
   \item Instruments = Cycling one by one
   \item Observation Mode = IMAG
   \item Data Type = OBJECT + COMPARISON + TEST
   \item Obs Category = SCIENCE + ENGINEERING
   \item Output format = ASCII
   \item Observation Date = Eventually split in intervals to avoid 10,000 lines limit 
   \item Output Columns = FRAMEID + DATE\_OBS + OBJECT + FILTER + RA2000 + DEC2000 + UT\_START + EXPTIME
\end{compactitem}
}

\subsection{Other Instrument Archives}

Additionally to major archive collections, we indexed two other instrument archives and 
surveys: \\

{\bf The Wide Field Imager (WFI) of the Anglo-Australian Telescope (AAT)} was ingested 
in the {\it Mega-Archive} in 2012. We considered important this instrument due to its large 
etendue, although this archive include only about 5,000 images observed in the period 
2000-2006. To manually collect the meta-data we used the former AAT Data Archive 
form\footnote{Old server http://site.aao.gov.au/arc-bin/wdb/aat\_database/observation\_log/make}
which was recently integrated in a modern portal serving more instrument 
archives\footnote{New server https://datacentral.org.au/archives}. 

\vspace{+1em}

{\bf The Sloan Digital Sky Survey (SDSS-III DS9 release)} was included in 2013, being 
queried via the DR9 Science Archive Server\footnote{https://dr9.sdss.org/fields}. 
The {\it Mega-Archive} includes almost one million images from the SDSS-III DR9 archive. 

\subsection{Migration to the SQL architecture}
\vspace{+1em}

The classic format of the {\it Mega-Archive} used between 2010 and 2017 was ASCII (txt files), 
each observation being stored in one line with columns defining data presented in Section~2.1
(separated by the ``\textbar'' character), and each instrument archive being stored in one file. \\

Following the migration to the actual EURONEAR dedicated server, in July 2017 the old 
{\it Mega-Archive} ASCII database was imported to SQL. This change was recommended to keep 
the search time as short as possible, due to the increase in number of images in {\it Mega-Archive}. 
We presented the architecture of {\it Mega-Archive} and its connection to the new EURONEAR 
tools as a flowchart in another recent paper \citep{Curelaru2019}. 
Thanks to the new SQL architecture, the searches of {\it Mega-Archive} have become faster for 
Solar system objects (taking a few minutes necessary to build accurate ephemerides needed to 
search the entire archive of 15 million observations for one asteroid) and extremely fast 
for fixed objects (taking only few fractions of a second for one target). 

\subsection{Mega-Archive Daily Update Crawler}
\vspace{+1em}

Major observatories ingest in their archives observations and images on a daily basis, which 
makes these databases appealing for searches related to very rapid time-domain phenomena. In 
other cases, dedicated surveys become ingested in archives much later, following the image 
reduction, project completion or expiration of proprietorship periods (typically one year 
following the observing date). 
Nevertheless, some newly discovered objects could rapidly raise attention of the astronomical 
communities, society and mass-media, such as a potential impactor asteroid, the interstellar object 
'Oumuamua or a close supernova. 
Sometimes such objects become very rapidly invisible even in largest telescopes, thus 
data mining remains the only possibility to prevent loss and obtain more information. 
In these cases, data mining of {\bf up to date} existing archives for precovery observations 
(preferably closer in time to discovery) could bring crucial information regarding the objects' 
nature, orbital classification and virtual or imminent impacts. \\

A {\it web crawler} (known also as a spider), is an Internet robot tool which systematically 
browses the World Wide Web with the aim to create entries to build or update a search engine 
index (e.g. \citep{Olston2010}). 
For automated update of the {\it Mega-Archive}, in 2014 we designed and implemented a crawler 
to check major archive collections on a daily basis. 
This PHP tool includes some scripts to query the ESO, CADC, SMOKA, ING and LCO collections, 
crawling the programmable interfaces published by each server (which very rarely need some minor 
changes of format). The only collection not able for automate crawling is the NVO archive, 
which we update manually on a yearly basis. 
The daily update tool starts to run automatically in cron every midnight and 
typically takes less than one hour to update all instrument archives in the {\it Mega-Archive}. 
Every SQL databases holding the instrumental archive could be updated (if there is new information) 
and also the master \texttt{ArchiveLogs.txt} file is updated (changing the number of images and the 
observed interval, needed for comparison with next day crawling). 
Each instrument archive is searched at once, checking only the most recent data (ingested in the 
collection during the previous day). Most problems (possible mal-functions due to servers or internet 
connection) could be traced and corrected next day by the admin who checks the logs listing the 
operations with all instrumental archives crawled in all collections, but in practice such system 
errors very rarely happen. 

\subsection{Archive Comparisons}
\vspace{+1em}

As of March 2019, the {\it Mega-Archive} includes 111 instruments, listed in Table~1. The columns list 
several parameters, namely the {\it archive etendue} ($A\Omega\,A$), the field of view of the instrument (FOV), 
the archival date interval, and the total number of raw science images indexed to date. Using these data, 
we could compare the archives and assess their use for data mining. \\

Figure~1 plots the histogram counting the total number of archives versus the number of images in each 
archive. The first three instruments are actually three surveys, namely CTIO1.3m (mostly 2MASS), VISTA 
and SDSS-DR9, followed by CTIO0.9m, INT-WFC and WHT-LIRIS. \\

Based on the etendue ($A\Omega$), the most powerful survey facilities in the present {\it Mega-Archive} 
are Subaru-HSC, Blanco-DECam, VISTA-VIRCAM, Subaru-SuprimeCam, VLT-VIMOS and CFHT-MegaCam. These and the 
following are plotted in the histogram in Figure~2. 
The analysis is extended in Figure~3 which plots the histogram counting the number of archives versus 
the {\it archive etendue} ($A\Omega\,A$). Based on this factor, the most productive facilities to date are 
VISTA-VIRCAM, Blanco-DECam, VST-OmegaCam, CFHT-MegaCam, Subaru-SuprimeCam, VLT-VIMOS and INT-WFC. Quite 
many near infrared (NIR) instruments populate Figures~1 and 3 due to their fast cadence of images explained 
by short exposure times and very rapid readouts compared to the other instruments in visible. \\

Most exposures are shorter than 80s, probably owing to the fast cadence of NIR instruments and other fast 
time-domain science, as one can observe the histogram in Figure~4. This makes {\it Mega-Archive} and data 
mining very appealing for asteroid science, as most targets should have stellar-like aspect (instead of 
longer trails) which results is easily measurable astrometry and photometry. \\

The number of images as a function of the observing time when they were obtained is shown in Figure~5. 
The great majority of the images in the actual {\it Mega-Archive} were collected since 2005. The last 
two years show lower numbers due to later ingestion in the collections and proprietorship which make 
last images invisible to collections and {\it Mega-Archive}. 

\section{Data Mining Tools}

{\it Mega-Archive} was designed in 2010 for asteroid searches and it was improved later, 
mostly after 2017 following the migration of EURONEAR to the new private dedicated server. 
We will present next the applications accessing the {\it Mega-Archive}. 
 
\subsection{Mega-Precovery for Moving Objects}
\vspace{+1em}

In 2010 we introduced {\it Mega-Precovery} to search for Solar System Objects using 
the {\it Mega-Archive} collection \citep{Popescu2010}. Our project targeted mostly the precovery 
and recovery of observations for PHAs and VIs. 
The algorithm and  a flowchart was presented in \citep{Vaduvescu2013}. 
During last years, several major improvements were added to the project, significantly extending 
its functionality. These are described bellow. \\

The original version (2010-2016) resided on the old EURONEAR IMCCE server. It was embedded in 
the old wikiplugin PHP environment and used the old ASCII archives format. The code migrated to 
the new EURONEAR server where it is preserved (under the Older Tools section) as version v.1 
\footnote{http://www.euronear.org/tools/megaprecoveryV1.php}. 
The second version (2016-2017) updated the code to pure PHP~7 environment installed in the new 
EURONEAR server, and converted the {\it Mega-Archive} to SQL database. The code has been 
preserved as version v.2\footnote{http://www.euronear.org/tools/megaprecoveryV2.php}. 
The actual version 3 (released in October 2017) has implemented two options for the calculus 
of the ephemerides of the searched object, allowing three options for the input which are offered 
as three different pages linked from the main EURONEAR Data Mining 
Tools\footnote{http://www.euronear.org/tools.php}. Moreover, additionally to the standard 
web form, the actual {\it Mega-Precovery} v.3 allows programmable queries via HTTP commands. 

\subsubsection{Mega-Precovery from Designations}
\vspace{+1em}

This is our first classic search for one or a few known asteroids or comets, based on 
the name, number or designation \footnote{http://www.euronear.org/tools/megaprecdes.php}. 
The user can chose between two ephemerides generators, either using the classic Miriade 
server\footnote{http://vo.imcce.fr/webservices/miriade} \citep{Berthier2009} (available in 
{\it Mega-Precovery} v.1, v.2 and v.3) or the {\it OrbFit}\footnote{http://adams.dm.unipi.it/orbfit} 
software \citep{OrbFit2011} installed only in {\it Mega-Precovery} v.3. 
The SsODNet service at IMCCE\footnote{http://vo.imcce.fr/webservices/ssodnet} 
\citep{Berthier2007} is used to check object designations. 

\subsubsection{Mega-Precovery from Orbit}
\vspace{+1em}

If the object doesn't have an official name, number or designation, but its orbital elements 
are know, then these can be input into Mega-Precovery\footnote{http://www.euronear.org/tools/megaprecorb.php}. 
The required elements are the semimajor axis $a$, the eccentricity $e$, the inclination $i$, the ascending 
node $\Omega$, the argument of periapsis $\omega$, the mean anomaly $M$, and the epoch $MJD$. 

\subsubsection{Mega-Precovery from Observations}
\vspace{+1em}

If the object was discovered very recently or if the connection with the ephemerides server 
is broken, then the target can be searched using an input consisting in a block of observations 
in MPC format\footnote{http://www.euronear.org/tools/megaprecobs.php}. In this case, {\it OrbFit} 
is run locally on the EURONEAR server, either using a single step ephemerides model (given at 
least 3 observations taken in one or few nights following discovery) or a three steps model 
(defined by blocks of discovery and follow-up data, first opposition data, and 
multi-opposition data). \\

All three search options are provided with a few advanced options: 
All instruments or only some archives selected by the user could be searched at once 
in {\it Mega-Precovery}. 
The computational interval can be constrained in an interval around the observations 
(to speed up the search of the entire {\it Mega-Archive}) or left default to cover all 
archives. 
The safety search border allows some flexibility due to telescope pointings (sometimes 
insecure, affected by small dithering during multiple run exposures, or due to the pointing 
not matching the camera centre). The default value of the safety border is $0.02^\circ$, 
which means that the search could allow any target outside the image by up to $1.2^\prime$ 
to be included in the results. 
The running mode allows the user to chose between fast geocentric ephemerides (where 
all ephemerides and steps are calculated at geocentre) and slow topocentric mode (where 
each integration step is calculated for the topocentric observatory position (recorded
as MPC code in \texttt{ArchiveLogs.txt} where the current searched instrument is hosted 
- this is important only for very close flyby observations of PHAs or VIs where 
topocentric correction could become important). 
Finally, three output options are provided: HTML, simple text (formatted as a table),
and CSV (fields separated by coma). \\

The {\it Mega-Precovery} output columns are: 
the archive name (in the archive/telescope-instrument format), 
the image ID (tentatively linked in HTML to the collection server to retrieve the FITS
image), the observed time (YYYY/MM/DD HH:MM:SS UT format), exposure time (in seconds), 
the expected object magnitude ($V$ when available), the telescope limiting magnitude 
(rough limit given the aperture, based on conventions mentioned in Section~2.1), 
filter and targeted object, angular distance to the field pointing (in degrees, 
followed by a percentage compared with the circular FOV), and a link to a plot 
showing the camera overlay (only for available mosaic cameras) marking the expected 
position of the searched object.

\subsection{MASFO for Fixed Objects}
\vspace{+1em}

In July 2017 we introduced MASFO\footnote{http://www.euronear.org/tools/masfo.php} tool
to allow {\it Mega-Archive Search for Fixed Objects} given in J2000 RA and DEC coordinates. 
One or a few objects (entered in successive lines) could be searched given the approximating $V$ 
magnitude of the object needed to compare with the limiting magnitude of the telescope 
(as given in Section~2). The output consists in the same columns as {\it Mega-Precovery}, 
including a link to the CCD overlay plot for most mosaic cameras included in {\it Mega-Archive}. 

\subsection{MASDS for Double Stars}
\vspace{+1em}

Following some data mining work in double stars using the OmegaCam archive 
\citep{Curelaru2017}, in 2018 we implemented some tools for data mining the entire 
{\it Mega-Archive} for known double stars given by the Washington Double Stars 
Catalog (WDS) ID or the WDS Discoverer ID. This tool is named {\it Mega Archive Search of 
Double Stars} (acronym MASDS)\footnote{http://www.euronear.org/tools/dstars/masds.php}. 
For observational planning we provide another tool named {\it WDS Filter 
Datamining}\footnote{http://www.euronear.org/tools/dstars/wdsfilter/wdsfilter.php} 
which allows as input the sky area (RA/DEC box limits), stars' magnitudes, separation, 
existing observations and discoverer. Both these tools are presented fully in a recent 
paper \citep{Curelaru2019}. 

\subsection{FindCCD for Moving and Fixed Objects}
\vspace{+1em}

Survey telescopes and larger field mosaic cameras are increasing in number, and 
{\it Mega-Archive} presently includes about 20 such powerful imaging instruments, 
with Subaru-HSC (104 CCDs) and Blanco-DECam (62 CCDs) leading the list. 
Ideally data mining tools need to point their users directly to the exact CCDs 
of such a mosaic camera possibly holding their fixed or mobile targets. \\

In 2013 we provided  
{\it Find CCD Subaru}\footnote{http://www.euronear.org/tools/findsubaruccd.php} 
to search for known NEAs in the SuprimeCam archive \citep{Vaduvescu2017}. In 2016 
we extended this work to other few cameras, namely VST-OmegaCam, INT-WFC, VISTA-VIRCAM, 
CFHT-MegaCam, Blanco-DECam and Subaru-HSC. 
This tool is named {\it FindCCD}\footnote{http://www.euronear.org/tools/findccdaster.php},
being developed for NEA searches. It queries the SkyBoT 
server\footnote{http://vo.imcce.fr/webservices/skybot} \citep{Berthier2006} to 
calculate positions of the known NEAs in the field and 
the NEODyS-2 server\footnote{https://newton.spacedys.com/neodys} \citep{Chesley1999} 
to overlay the uncertainty ellipses for poorly observed NEAs. In Figure~6 we give an 
example of {\it FindCCD} plot for the very poorly observed NEA 2015~BS516 searched in a 
given Blanco-DECam image, where the uncertainty region (plotted in red) covers many CCDs. \\

In March 2018 we released
{\it FindCCD for Double Stars}\footnote{http://www.euronear.org/tools/dstars/findccddstars.php} 
to identify the particular CCD of few major mosaic cameras possible to hold known double stars 
from the Washington Star Catalog \citep{Curelaru2019}. \\

In June 2018 we deployed {\it FindCCD for 
Fixed Objects}\footnote{http://www.euronear.org/tools/findccdfixed.php}
to search for stars, galaxies or other fixed objects given as J2000 RA/DEC coordinates. 
Seven mosaic cameras are supported by March 2019, but we plan to extend the list soon. 
Figure~7 gives the result of the search of a galaxy in one VIRCAM image.

\section{Conclusions and Recommendations}

The EURONEAR {\it Mega-Archive} project started in 2010 by joining meta-data of the first 
three instrument archives (CFHTLS, ESO/MPG and ING/INT), becoming the first such server for 
data mining asteroids and NEAs. 
In the same year we introduced {\it Mega-Precovery} to search {\it Mega-Archive} images for 
one or a few known asteroids or comets given by designations. \\

During past years we added other archives, mainly based on six archive collections (CADC, 
ESO, ING, SMOKA, NVO and LCOGT) and by 2018 {\it Mega-Archive} has indexed 15 million images. 
In 2014 we implemented a crawler for automate query of five collections (except for NVO 
which does not allow programmatic queries) for daily update of the {\it Mega-Archive}. 
In 2016 we released the second version of {\it Mega-Archive} and {\it Mega-Precovery} 
which migrated to the new EURONEAR server and adopted the new SQL database architecture. 
Since Oct 2017 the third actual version of {\it Mega-Precovery} introduced two options 
for calculus of the ephemerides and three input options (search by designation, based on 
orbit, and observations). \\

During last few years we designed other tools aimed to take advantage of {\it Mega-Archive}
for science other than NEAs. 
In 2017 we introduced MASFO tool to search the {\it Mega-Archive for Fixed Objects}. 
In 2018 we implemented the MASDS tool for {\it Mega-Archive Search for Double Stars}. 
To specify exactly the exact CCD of major mosaic cameras include in the {\it Mega-Archive}, 
in 2016 and 2018 we built the graphical {\it FindCCD} and {\it FindCCD for Fixed Objects}
to overlay the moving or fixed targets over seven mosaic cameras, plotting also the uncertainty 
ellipse for poorly observed NEAs. 
In the near future we plan to grow the {\it Mega-Archive} and improve {\it Mega-Precovery} 
and other related data mining tools. \\

We have entered already in the big data epoch, where present and future surveys will 
provide huge amount of imaging data valuable for data mining and time-domain astronomy. 
In this sense, we recommend to the IAU, specifically Division B (Facilities, Technologies 
and Data Science) and Commissions B2 (Data and Documentation) and B3 (Astroinformatics 
and Astrostatistics) to adopt a common format and recommend to all astronomical 
observatories to use it for indexing and storing their science images, and make them 
available for programmable queries. \\

On a cosmic scale, NEAs pose real threat to mankind, and there are cases when faint 
or/and very fast moving virtual or imminent impactors could not be recovered even by 
the largest telescopes which typically are lacking immediate observing time. 
In such cases, tools able to datamine very recent archives could become crucial to 
precover such objects, improve their orbits, assess the risks and eventually eliminate 
the impact threats. 
In this sense, we recommend to astronomical observatories to index their entire 
available observations (completely up to to date) even though some images are considered 
the proprietorship of a given PI or project (who could decide to collaborate per request 
in making available such images whenever such scenario will raise). \\

\section{Acknowledgements}

This research used the facilities of the Canadian Astronomy Data Centre operated by 
the National Research Council of Canada with the support of the Canadian Space Agency. 
This work is based on data obtained from the ESO Science Archive Facility. 
The SMOKA archive is operated by the Astronomy Data Center and the National Astronomical 
Observatory of Japan, being based on data collected at Subaru Telescope, Okayama Astrophysical 
Observatory, Kiso observatory (University of Tokyo) and Higashi-Hiroshima Observatory. 
This research uses services or data provided by the Science Data Archive at NOAO. 
NOAO is operated by the Association of Universities for Research in Astronomy (AURA), Inc. 
under a cooperative agreement with the National Science Foundation. The first manual search 
of the archive was performed by the EURONEAR collaborator Farid Char (University of 
Antofagasta, Chile). 
This paper makes use of data obtained from the Isaac Newton Group of Telescopes Archive 
which is maintained as part of the CASU Astronomical Data Centre at the Institute of Astronomy, 
Cambridge. The ING archive was built from the observing logs during a summer project by the 
student Vlad Tudor (former ING student and EURONEAR collaborator). 
This work makes use of the Science archive and observations from the LCOGT network. The first 
archive search and retrieval was performed by the EURONEAR collaborators Ioana and Adrian Stelea
(Bucharest Astroclub, Romania). 
MASDS for Double Stars tool has made use of the Washington Double Star catalogs maintained 
at the U.S. Naval Observatory. 
This research has made use of Miriade, SkyBoT and SsODNet VO servers developed at IMCCE, 
{\it Observatoire de Paris}. We thank to Jerome Berthier for his continuous support regarding 
the access to these services.  
MP acknowledges support from the AYA2015-67772-R (MINECO, Spain).

\newpage
\begin{longtable}{lrrrrr}
\caption{15 million images from 111 instrument archives indexed in the {\it Mega-Archive} by 23 Feb 2019. } \\
\hline\hline\noalign{\smallskip}
Instrument Archive      & $A\Omega\,A$ &  FOV  &   Start Date &     End Date &  Imags  \\
\noalign{\smallskip}\hline\hline\noalign{\smallskip}
\endfirsthead
\caption{continued.}\\
\hline\hline\noalign{\smallskip}
Instrument Archive      & $A\Omega\,A$ &  FOV  &   Start Date &     End Date &  Imags  \\
\noalign{\smallskip}\hline\hline\noalign{\smallskip}
\endhead
\noalign{\smallskip}\hline\hline\noalign{\smallskip}
\endfoot
\noalign{\smallskip}
AAT-WFI                    &     14831 &  1089.0 &  2000-08-21  &  2006-02-05  &     4453 \\
\noalign{\smallskip}\noalign{\smallskip}                                                                                          
CADC/APASS                 &     42242 & 31329.0 &  2010-04-11  &  2014-06-02  &   121350 \\
CADC/CFHT-aobir            &       184 &     1.4 &  1997-12-10  &  2011-11-23  &    45108 \\
CADC/CFHT-aobvis           &         6 &     1.4 &  1996-05-04  &  1999-03-30  &     1573 \\
CADC/CFHT-CFHTIR           &      1397 &    17.6 &  2001-01-10  &  2005-11-17  &    27996 \\
CADC/CFHT-Megacam          &   1464861 &  3600.0 &  2003-02-22  &  2013-05-09  &   143896 \\
CADC/CFHT-MOCAM            &       421 &   225.0 &  1995-05-28  &  1995-11-29  &      662 \\
CADC/CFHT-REDEYE           &       241 &     4.4 &  1993-02-04  &  1998-09-03  &    19312 \\
CADC/CFHT-WIRCam           &    437812 &   466.6 &  2005-11-18  &  2019-01-23  &   331845 \\
CADC/CTIO-CPAPIR           &     19191 &   348.2 &  2005-02-13  &  2019-02-15  &   112290 \\
CADC/CTIO-CPAPIRVIS        &      1616 &   348.2 &  2008-01-16  &  2018-10-27  &     9453 \\
CADC/DAO-E2VCCD            &      9125 &   253.6 &  2008-02-11  &  2019-03-04  &    52018 \\
CADC/GeminiN-GNIRS         &       190 &     1.4 &  2004-09-05  &  2015-03-15  &     9232 \\
CADC/GeminiN-GMOS          &     12200 &    30.5 &  2001-08-14  &  2015-12-01  &    27973 \\
CADC/GeminiN-NIRINIR       &     10436 &     3.2 &  2002-02-22  &  2015-12-02  &   225033 \\
CADC/GeminiN-NIRIVIS       &       101 &     3.2 &  2004-06-28  &  2015-10-07  &     2181 \\
CADC/GeminiS-Flamingos2    &      6439 &    36.0 &  2013-10-10  &  2015-11-29  &    12496 \\
CADC/GeminiS-GMOS          &     13299 &    30.5 &  2003-02-28  &  2015-11-29  &    30491 \\
CADC/HST-ACS               &      1277 &    11.7 &  2002-04-02  &  2018-03-19  &    88290 \\
CADC/HST-NICMOS            &       114 &     0.8 &  1997-03-21  &  2008-09-10  &   113867 \\
CADC/HST-NICMOS            &         2 &     0.8 &  1997-07-23  &  2008-08-21  &     1730 \\
CADC/HST-WFC3/NIR          &       186 &     4.9 &  2009-07-05  &  2012-11-15  &    30543 \\
CADC/HST-WFC3/Vis          &       186 &     7.3 &  2009-07-13  &  2012-11-15  &    20626 \\
CADC/HST-WFPC/OPT          &        41 &     9.0 &  1989-11-30  &  1993-12-03  &     3691 \\
CADC/HST-WFPC2/NIR         &        15 &     9.0 &  1994-03-05  &  2008-04-06  &     1366 \\
CADC/HST-WFPC2/OPT         &      1803 &     9.0 &  1993-12-20  &  2009-05-12  &   162072 \\
CADC/Mt.Stromlo-MACHO      &    114404 &  1794.4 &  1992-07-21  &  2002-11-19  &   197868 \\
CADC/UKIRT-Michelle        &       160 &     1.4 &  2001-10-04  &  2004-04-23  &    35337 \\
CADC/UKIRT-UFTINIR         &      1871 &     3.2 &  1999-10-16  &  2011-07-18  &   183298 \\
\noalign{\smallskip}\noalign{\smallskip}
ESO/3.6m-TIMMI2            &       305 &     1.9 &  2004-05-08  &  2006-06-28  &    63817 \\
ESO/MPG-WFI                &    144865 &  1149.1 &  1999-04-15  &  2018-03-22  &   139642 \\
ESO/NTT-EMMI               &      3834 &    83.2 &  2004-03-17  &  2008-04-01  &    17541 \\
ESO/NTT-SOFI               &     17201 &    24.2 &  2006-03-30  &  2018-03-14  &   270421 \\
ESO/NTT-SUSI2              &      1366 &    30.5 &  2004-04-02  &  2008-12-29  &    17057 \\
ESO/VISTA-VIRCAM           &  22270114 &  4726.6 &  2009-10-16  &  2018-03-23  &  1414673 \\
ESO/VLT-EFOSC2             &     24864 &    16.7 &  2004-07-03  &  2018-03-12  &   103169 \\
ESO/VLT-FORS1              &     23860 &    46.0 &  1999-01-23  &  2009-03-26  &    35852 \\
ESO/VLT-FORS2              &    129969 &    46.0 &  1999-10-30  &  2018-03-21  &   195289 \\
ESO/VLT-HAWKI              &    105972 &    56.3 &  2007-08-01  &  2018-03-22  &   130127 \\
ESO/VLT-ISAAC              &     19153 &     6.4 &  1999-03-01  &  2013-12-13  &   208325 \\
ESO/VLT-NACO               &      4719 &     0.9 &  2001-12-02  &  2018-03-22  &   353652 \\
ESO/VLT-VIMOS              &   1134094 &   841.8 &  2002-10-30  &  2018-02-15  &    93060 \\
ESO/VLT-VISIR              &       137 &     0.1 &  2004-05-11  &  2015-11-22  &    73068 \\
ESO/VST-OMEGACAM           &   1142071 &  3664.5 &  2011-04-01  &  2018-03-23  &   240767 \\
\noalign{\smallskip}\noalign{\smallskip}                                                                                         
ING/JKT-JAG                &      2158 &   100.4 &  2002-01-03  &  2003-08-01  &    98553 \\ 
ING/WHT-ACAM               &     24856 &    63.7 &  2009-06-10  &  2018-03-05  &   101428 \\
ING/WHT-LDSS               &      1276 &   100.4 &  1993-03-18  &  2000-03-24  &     3302 \\
ING/WHT-LIRIS              &     39347 &    18.2 &  2004-03-03  &  2018-01-29  &   563405 \\
ING/WHT-PFIP               &     20328 &   256.6 &  1993-12-03  &  2016-09-22  &    20582 \\
ING/WHT-WHIRCAM            &        13 &     1.0 &  1995-02-10  &  1999-06-28  &     3174 \\
ING/INT-PFCCD              &       221 &   121.0 &  1993-01-13  &  2014-05-17  &     1492 \\
ING/INT-WFC                &    922193 &  1169.6 &  2002-01-02  &  2018-03-05  &   643626 \\
\noalign{\smallskip}\noalign{\smallskip}  
LCOGT/SSO-0M4-03           &       581 &   582.1 &  2015-10-15  &  2019-03-04  &    29931 \\
LCOGT/SSO-0M4-05           &       424 &   582.1 &  2015-10-02  &  2019-03-04  &    21830 \\
LCOGT/SSO-1M0-03           &      3482 &   169.5 &  2014-05-01  &  2019-03-04  &    94788 \\
LCOGT/SSO-1M0-11           &      2370 &   169.5 &  2014-05-01  &  2019-03-04  &    62911 \\
LCOGT/SSO-2M0-02           &      3016 &   100.4 &  2014-05-01  &  2019-03-04  &    34442 \\
LCOGT/HLK-0M4-04           &       206 &   582.1 &  2016-03-31  &  2019-03-04  &    10633 \\
LCOGT/HLK-0M4-06           &      1417 &   582.1 &  2016-03-31  &  2019-03-04  &    73010 \\
LCOGT/HLK-2M0-01           &      5386 &   100.4 &  2014-05-04  &  2019-03-04  &    61509 \\
LCOGT/CTIO-0M4-09          &       222 &   582.1 &  2017-12-04  &  2019-03-05  &    11432 \\
LCOGT/CTIO-1M0-04          &      1629 &   697.0 &  2014-05-12  &  2019-03-05  &    10790 \\
LCOGT/CTIO-1M0-05          &     15907 &   697.0 &  2014-06-01  &  2019-03-05  &   105336 \\
LCOGT/CTIO-1M0-09          &       793 &   697.0 &  2014-05-31  &  2019-03-05  &     5251 \\
LCOGT/OT-0M4-10            &       266 &   582.1 &  2015-06-03  &  2019-03-05  &    12645 \\
LCOGT/OT-0M4-14            &      1851 &   582.1 &  2015-02-13  &  2019-03-05  &    88046 \\
LCOGT/OT-0M4-XX            &        54 &   582.1 &  2015-02-15  &  2015-06-02  &     2577 \\
LCOGT/SED-0M8-01           &       513 &   125.3 &  2014-05-02  &  2018-02-14  &    36883 \\
LCOGT/GOL-1M0-02           &       213 &   169.5 &  2014-05-02  &  2018-08-23  &     5795 \\
LCOGT/SAAO-1M0-10          &     17436 &   697.0 &  2014-05-01  &  2019-03-04  &   112576 \\
LCOGT/SAAO-1M0-12          &     10098 &   697.0 &  2014-05-01  &  2019-03-04  &    65198 \\
LCOGT/SAAO-1M0-13          &     15424 &   697.0 &  2014-05-01  &  2019-03-04  &    99589 \\
LCOGT/MDO-1M0-08           &     10158 &   697.0 &  2014-05-02  &  2019-03-04  &    67269 \\                                                                                        
\noalign{\smallskip}\noalign{\smallskip}
NVO/CTIO0.9m-CCD           &     24679 &   182.3 &  2000-04-04  &  2017-05-17  &   826261 \\
NVO/CTIO1m-Y4KCam          &     22409 &   392.0 &  2006-12-14  &  2014-06-02  &   263817 \\
NVO/CTIO1.3m-ANDICAM       &      4524 &     5.8 &  2000-04-04  &  2017-06-01  &  2141898 \\
NVO/Blanco-MOSAIC2         &    322956 &  1388.3 &  2004-08-11  &  2012-02-20  &    83665 \\
NVO/Blanco-NEWFIRM         &    126046 &   785.1 &  2010-07-03  &  2011-05-19  &    57738 \\
NVO/Blanco-DECam           &  15142160 & 17424.0 &  2012-09-12  &  2017-06-05  &   312542 \\
NVO/Blanco-ISPI            &     30117 &   104.0 &  2006-06-05  &  2014-04-10  &   104106 \\
NVO/Bok-CCD                &    197417 &  4844.2 &  2015-01-08  &  2017-06-04  &    35610 \\
NVO/KPNO2.1m-CCD           &       859 &    29.2 &  2010-09-03  &  2014-07-02  &    35940 \\
NVO/Mayall-Misc            &        64 &     0.4 &  2007-02-21  &  2015-06-29  &    70005 \\
NVO/Mayall-MOSAIC          &    109052 &  1296.0 &  2004-09-01  &  2013-02-07  &    32998 \\
NVO/Mayall-NEWFIRM         &    260610 &   785.1 &  2007-06-30  &  2013-02-04  &   130171 \\
NVO/SOAR-SAM               &       581 &     9.0 &  2014-03-05  &  2017-05-23  &    18689 \\
NVO/SOAR-OptImg            &     19150 &    27.3 &  2005-11-07  &  2017-06-14  &   203377 \\
NVO/SOAR-Spartan           &      7929 &    25.4 &  2011-08-16  &  2017-05-06  &    90331 \\
NVO/WIYN0.9m-MOSAIC        &     11534 &  3478.6 &  2004-11-25  &  2010-03-29  &    20232 \\
NVO/WIYN0.9m-S2KB          &     11705 &   416.2 &  2004-09-09  &  2014-11-17  &   171614 \\
NVO/WIYN3.5m-MiniMosaic    &      2887 &   100.4 &  2004-12-18  &  2013-04-20  &    12989 \\
NVO/WIYN3.5m-WHIRC         &      6009 &    10.9 &  2008-04-15  &  2017-05-19  &   249238 \\
\noalign{\smallskip}\noalign{\smallskip}                                                                                             
SDSS-DR9                   &    131306 &   140.8 &  1998-09-19  &  2009-11-18  &   938046 \\
\noalign{\smallskip}\noalign{\smallskip}
SMOKA/KANATA-HONIR         &      2075 &   100.4 &  2014-03-05  &  2016-10-30  &    42098 \\
SMOKA/KANATA-HOWPol        &     11803 &   225.0 &  2008-12-06  &  2016-10-24  &   106875 \\
SMOKA/Kiso-1kCCD           &      1123 &   155.8 &  1993-02-26  &  2000-10-24  &    30004 \\
SMOKA/Kiso-2kCCD           &     70609 &  2498.0 &  1998-09-08  &  2012-02-27  &   117640 \\
SMOKA/Kiso-KWFC            &    373092 & 15620.0 &  2012-04-02  &  2017-04-30  &    99408 \\
SMOKA/Okayama-ISLE         &      1146 &    17.1 &  2006-11-15  &  2016-04-24  &    86736 \\
SMOKA/Okayama-KOOLS        &        64 &    21.8 &  2008-01-04  &  2016-01-10  &     3822 \\
SMOKA/Okayama-OASIS        &        85 &    16.2 &  1998-08-14  &  1998-12-13  &     6843 \\
SMOKA/Subaru-CAC           &        45 &     0.7 &  1999-01-06  &  2000-06-21  &     4318 \\
SMOKA/Subaru-COMICS        &       433 &     0.5 &  1999-12-14  &  2016-07-26  &    56943 \\
SMOKA/Subaru-CISCO         &      8046 &     3.7 &  1999-01-12  &  2007-04-09  &   148791 \\
SMOKA/Subaru-FOCAS         &      6040 &    36.0 &  2000-02-02  &  2016-10-11  &    11437 \\
SMOKA/Subaru-HSC           &    129398 &  8100.0 &  2014-03-26  &  2014-07-08  &     1089 \\
SMOKA/Subaru-ICRS          &      2274 &     1.0 &  2000-09-22  &  2016-09-26  &   148973 \\
SMOKA/Subaru-Kyoto         &        61 &     3.9 &  2004-04-08  &  2015-09-24  &     1065 \\
SMOKA/Subaru-MOIRCS        &     30417 &    28.2 &  2004-06-11  &  2016-10-31  &    73474 \\
SMOKA/Subaru-SuprimeCam    &   1133632 &   918.5 &  1999-01-05  &  2016-05-09  &    92447 \\
SMOKA/Subaru-SCExAO        &       405 &     0.2 &  2000-01-22  &  2008-07-15  &   119955 \\

\end{longtable}


\bibliographystyle{elsarticle-num}
\bibliography{paper114}


\clearpage
\begin{figure}
\centering
  \includegraphics[width=14cm]{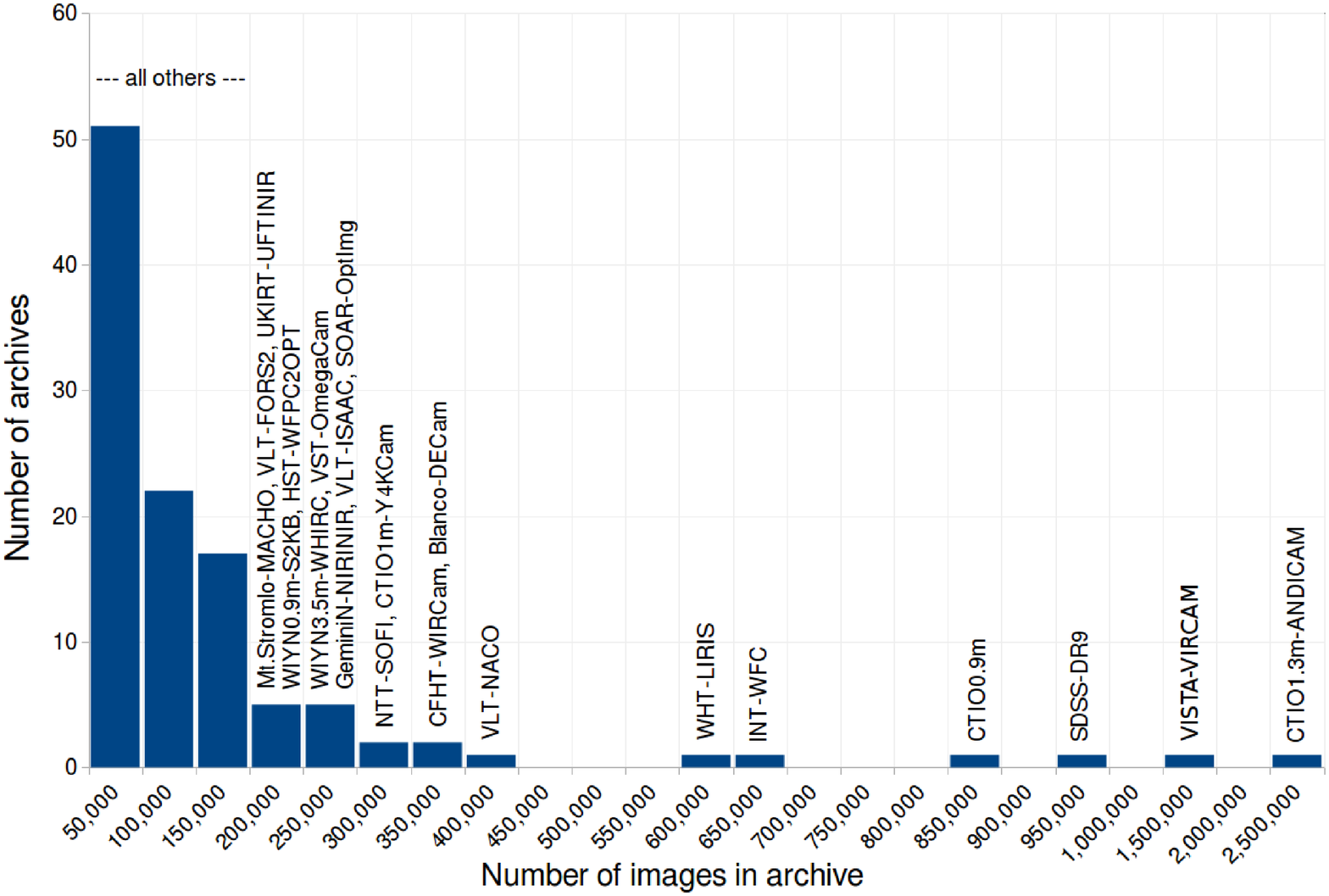} 
\begin{center}
{\bf Figure 1: } The number of archives versus the number of images in the {\it Mega-Archive}. 
Only the most populated instruments are labeled, and the step of the $X$ axis is changed above 
$X=1,000,000$ to be able to accommodate the most prolific two archives. 
\end{center}
\end{figure}

\begin{figure}
\centering
  \includegraphics[width=14cm]{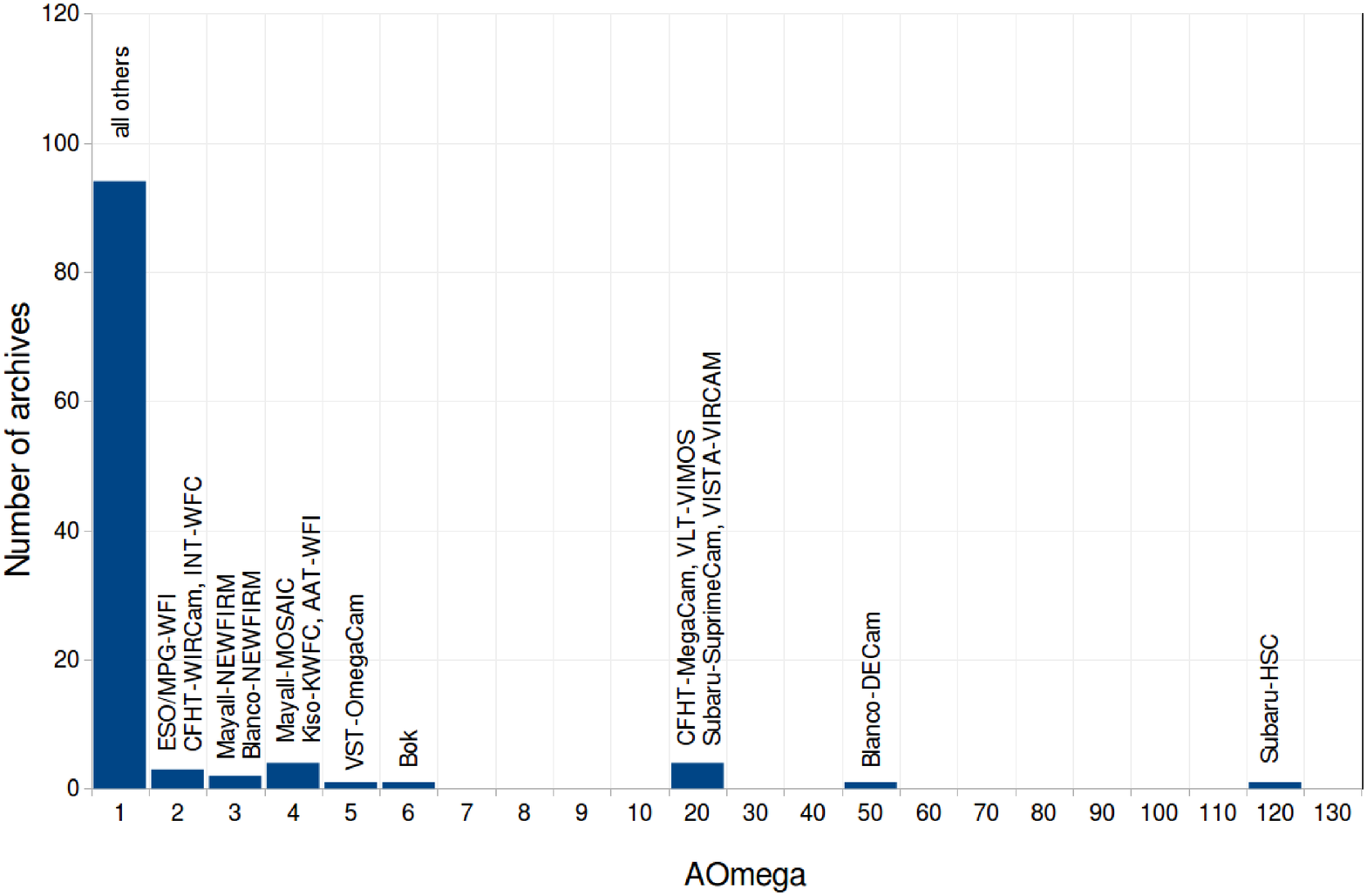} 
\begin{center}
{\bf Figure 2: } The number of archives versus the etendue ($A\Omega$) in the {\it Mega-Archive}. 
Only the most powerful instruments are labeled and the step on the $X$ axis is changed above $A\Omega=10$. 
\end{center}
\end{figure}

\begin{figure}
\centering
  \includegraphics[width=14cm]{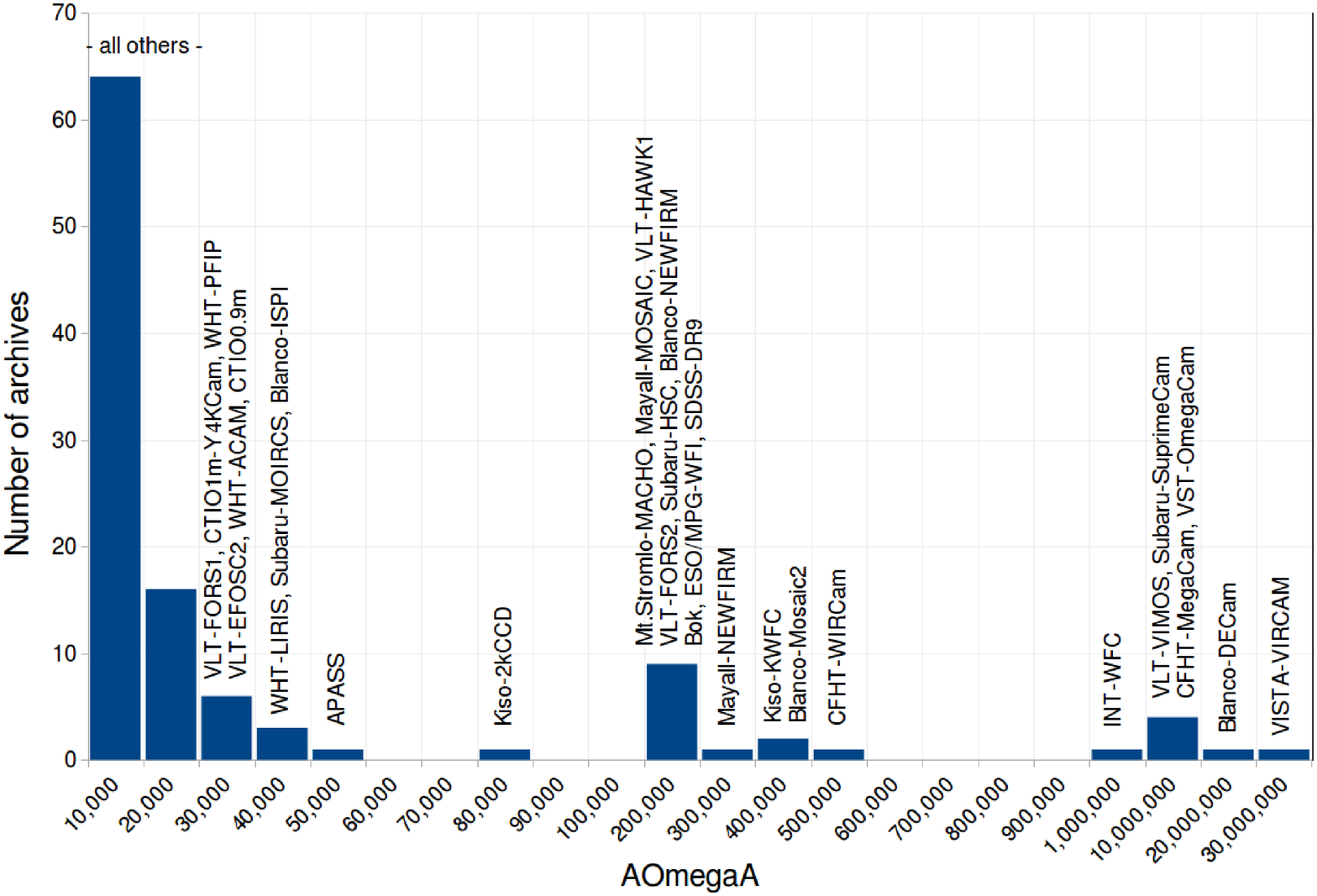} 
\begin{center}
{\bf Figure 3: } The number of archives versus the {\it archive etendue} ($A\Omega\,A$) in the {\it Mega-Archive}. 
Only the most powerful instrument archives are labeled across the variable step for the $X$ axis (changed at
$X=100.000$ and $X=1,000,000$), to be able to accommodate the most prolific archives. 
\end{center}
\end{figure}

\begin{figure}
\centering
  \includegraphics[width=14cm]{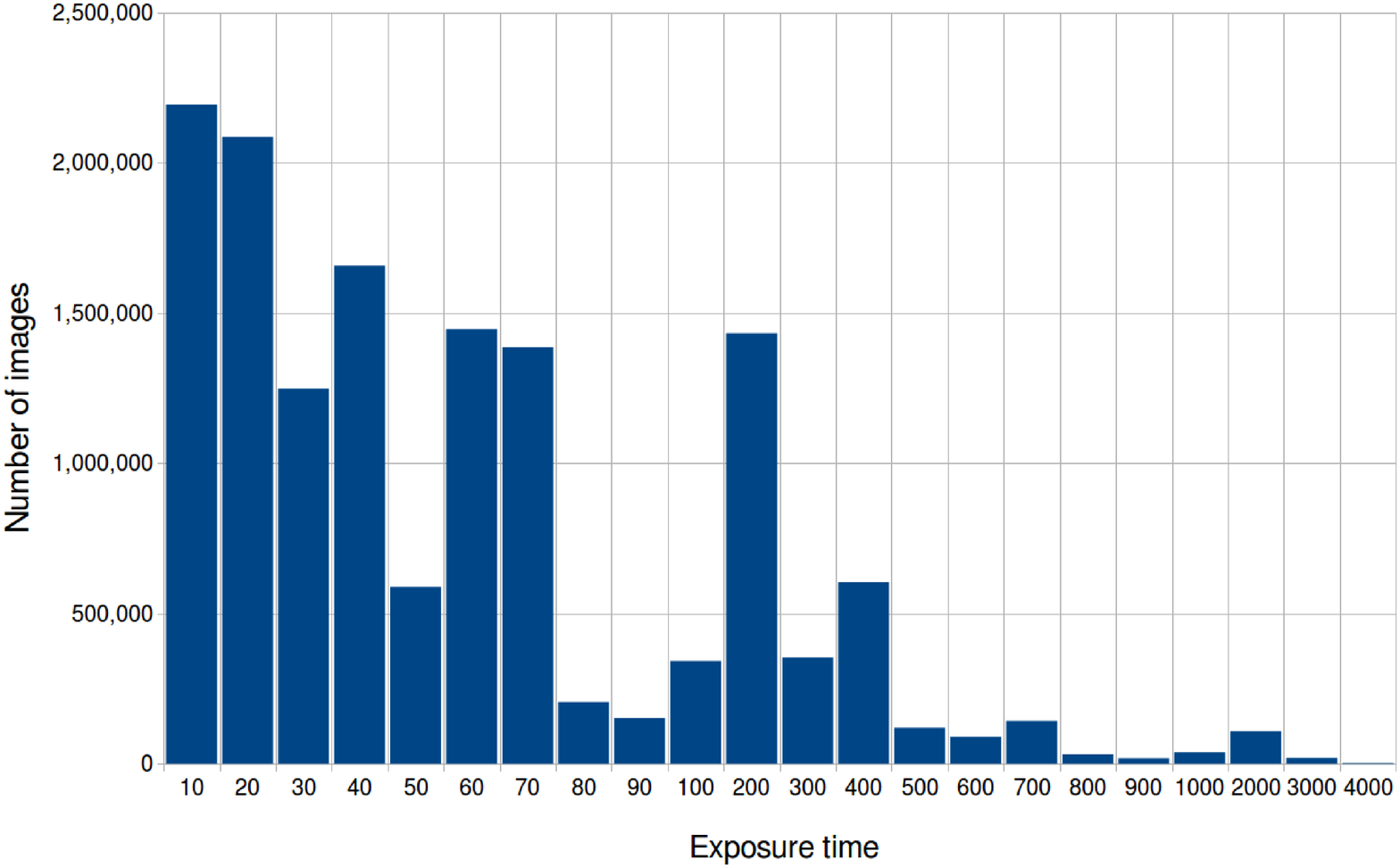} 
\begin{center}
{\bf Figure 4: } The number of images versus the exposure time (in seconds) in the {\it Mega-Archive}. 
The step on the $X$ axis is one unit larger for the last few cuts, to be able to accommodate all exposures. 
\end{center}
\end{figure}

\begin{figure}
\centering
  \includegraphics[width=14cm]{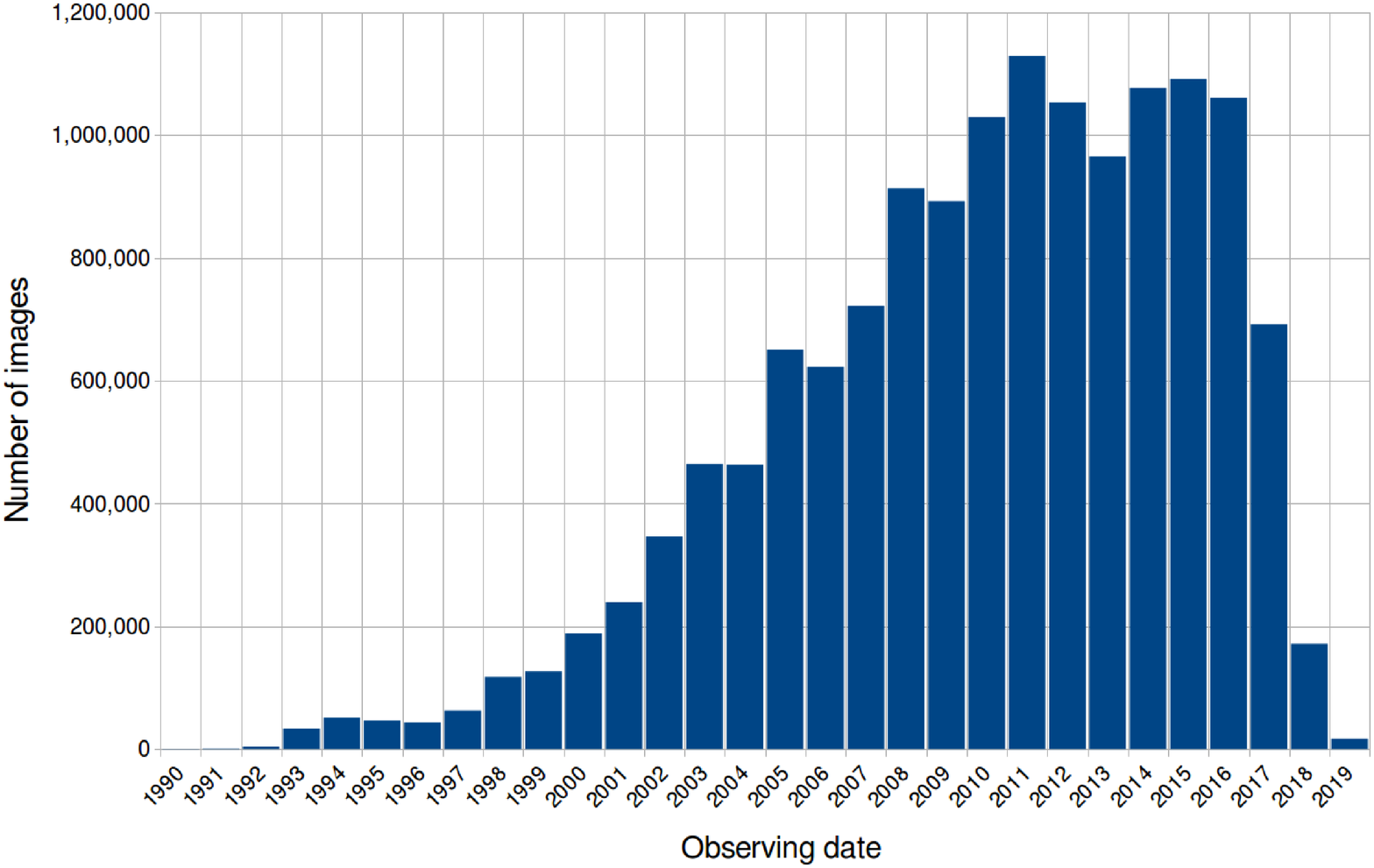} 
\begin{center}
{\bf Figure 5: } The number of images indexed in the {\it Mega-Archive} versus the observing date (year). 
\end{center}
\end{figure}

\begin{figure}
\centering
  \fbox{ \includegraphics[width=10cm]{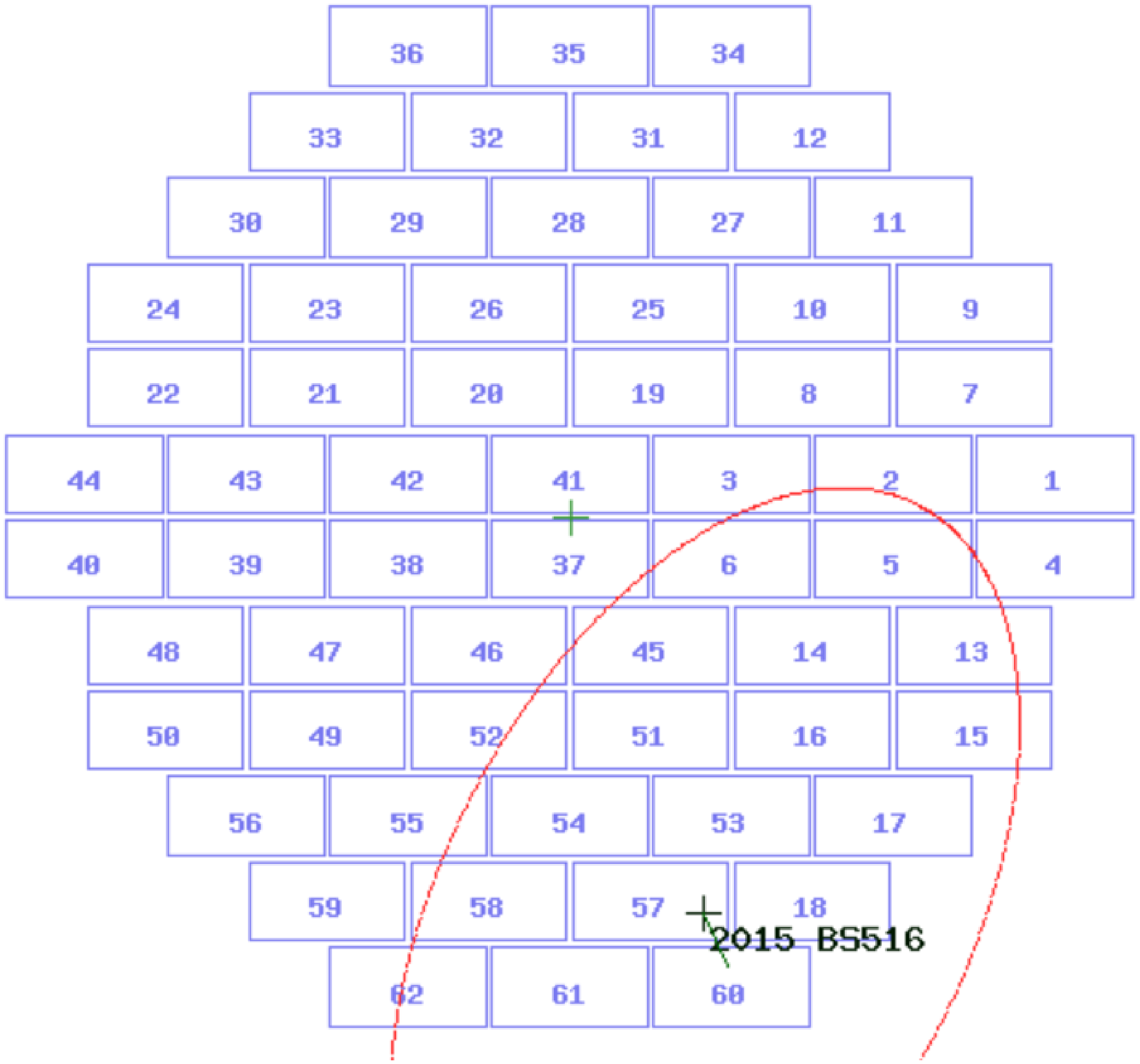} }
\begin{center}
{\bf Figure 6: } {\it FindCCD} plot showing the Blanco-DECam overlay of the image 
\texttt{c4d150213{\_}085354{\_}ooi{\_}g{\_}v1} and the uncertainty position of the poorly observed 
NEA 2015~BS516 (bordered in red) covering many CCDs.
\end{center}
\end{figure}

\begin{figure}
\centering
  \fbox{ \includegraphics[width=10cm]{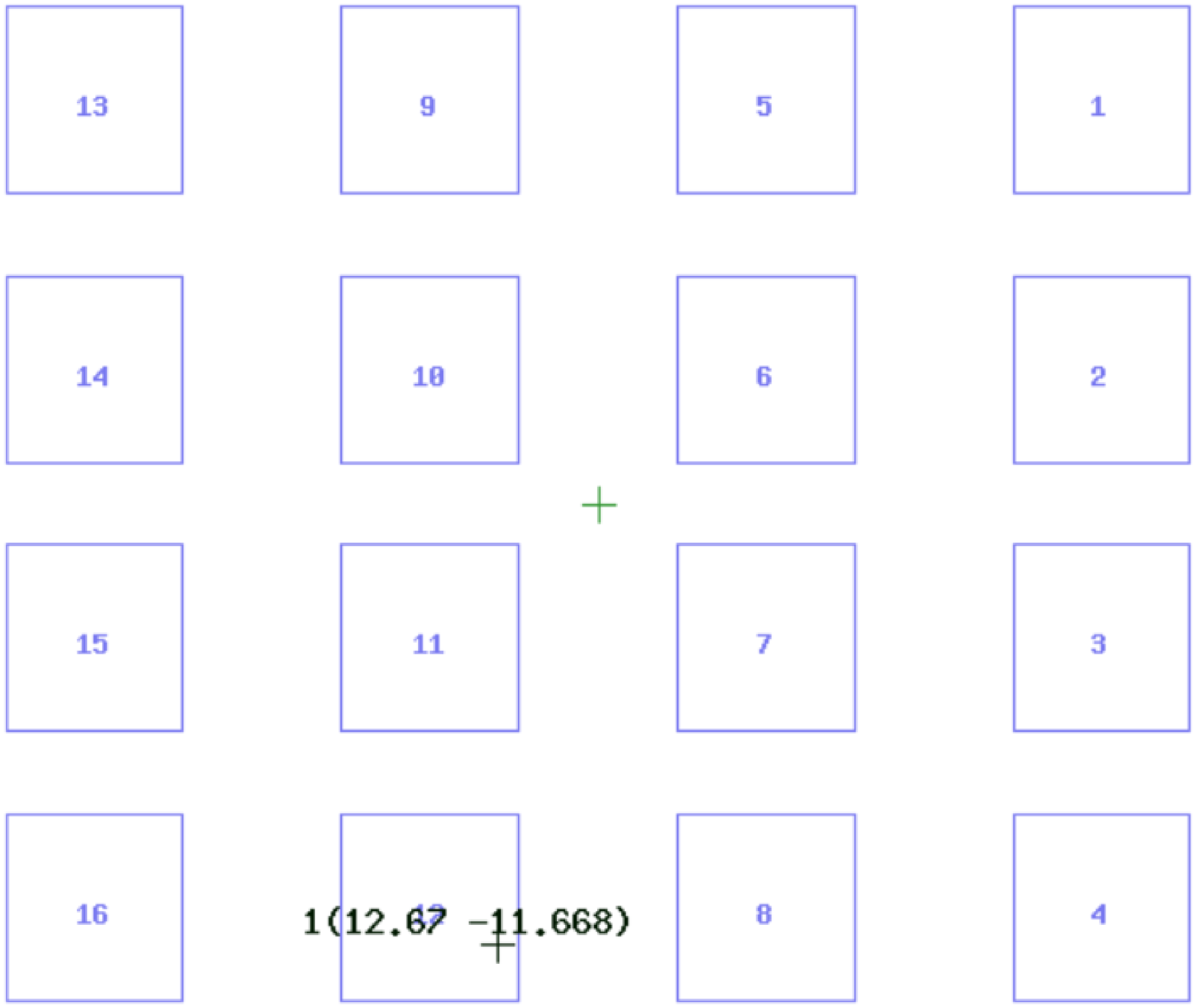} }
\begin{center}
{\bf Figure 7: } {\it FindCCD for Fixed Objects} plot showing the CCD holding a galaxy of coordinates 
12:40:03.1 -11:40:04 overlaid on the VISTA-VIRCAM mosaic image \texttt{VCAM.2013-01-21T07:52:15.676}.  
\end{center}
\end{figure}
 

\end{document}